\documentclass[runningheads]{llncs}
\usepackage[acronym, shortcuts]{glossaries-extra}
\usepackage{graphicx}
\usepackage[hidelinks]{hyperref}
\usepackage{multirow}
\usepackage{tikz}
\usepackage[moderate, tracking=normal, leading=normal]{savetrees}
\usepackage{subcaption}
\usepackage{soul}
\usepackage{xcolor}

\usepackage[strict]{changepage}

\usepackage{framed}
\definecolor{formalshade}{rgb}{0.85,1,0.85}
\definecolor{darkblue}{rgb}{0.0,0.6,0.30}

\usepackage{tikz}

\usepackage{pifont}

\begin{document}
%
\title{GAN-GRID: A Novel Generative  Attack on Smart Grid Stability Prediction}
\titlerunning{GAN-GRID}
%
\author{Emad~Efatinasab\inst{1} \and
Alessandro~Brighente\inst{2} \and
Mirco~Rampazzo\inst{1} \and
Nahal~Azadi\inst{1} \and
Mauro~Conti\inst{2, 3}}
\authorrunning{E. Efatinasab et al.}
%
\institute{University of Padova, Department of Information Engineering \and
University of Padova, Department of Mathematics \and
Delft~University~of~Technology, Faculty of Electrical Engineering, Mathematics and Computer Science \\
\email{\{emad.efatinasab@phd.unipd.it,nahal.azadi@studenti.unipd.it}
\email{\{alessandro.brighente, mirco.rampazzo, mauro.conti\}@unipd.it}}
\maketitle              
\begin{abstract}
The smart grid represents a pivotal innovation in modernizing the electricity sector, offering an intelligent, digitalized energy network capable of optimizing energy delivery from source to consumer.
It hence represents the backbone of the energy sector of a nation.
Due to its central role, the availability of the smart grid is paramount and is hence necessary to have in-depth control of its operations and safety.
To this aim, researchers developed multiple solutions to assess the smart grid's stability and guarantee that it operates in a safe state.
Artificial intelligence and Machine learning algorithms have proven to be effective measures to accurately predict the smart grid's stability.
Despite the presence of known adversarial attacks and potential solutions, currently, there exists no standardized measure to protect smart grids against this threat, leaving them open to new adversarial attacks.


In this paper, we propose GAN-GRID a novel adversarial attack targeting the stability prediction system of a smart grid tailored to real-world constraints. 
Our findings reveal that an adversary armed solely with the stability model's output, devoid of data or model knowledge, can craft data classified as stable with an Attack Success Rate (ASR) of 0.99. 
Also by manipulating authentic data and sensor values, the attacker can amplify grid issues, potentially undetected due to a compromised stability prediction system. These results underscore the imperative of fortifying smart grid security mechanisms against adversarial manipulation to uphold system stability and reliability.

\end{abstract}

\section{Introduction}
Smart Grid (SG) technology represents a modern electric power grid characterized by increased reliability, efficiency, sustainability, and bi-directional communication capabilities~\cite{en16010472}. 
By integrating advanced hardware (such as phasor measurement units and smart meters) and advanced software solutions, SGs provide safety and stability while concurrently reducing operational costs compared to previous energy distribution systems~\cite{7926429}. 
With the current urge to include renewable energy sources in the power market, the SG should be open to seamlessly including novel technologies together with their operations characteristics in terms of when they collect power, and how much power they can deliver.
Accurately predicting renewable energy generation is crucial for ensuring the stable, efficient, and cost-effective operation of the power system~\cite{Jiao_2020}.
To this aim, researchers developed stability prediction systems as software components of the SG.
They aim to collect data from the grid and, via historical analysis, predict whether the current configuration of the SG is going towards an unstable state to reconfigure it and guarantee service availability.
Machine Learning (ML) and Artificial Intelligence (AI) have proved to be a very efficient solution to this aim, with researchers proposing many different models with very good performance~\cite{onder2023classification,https://doi.org/10.1002/2050-7038.12706,breviglieri2021predicting,en13102559,sym15020289,DEWANGAN2022149,10.1145/3459104.3459160,9079864,9446196}

SGs represent the energy backbone of a nation and are hence among the critical infrastructures to be protected~\cite{forbes}.
Indeed, critical infrastructures have been recently targets of many cyber attacks, as their disruption might significantly impact a whole country~\cite{colonial}.
Several factors contribute to SG's attack surface and hence vulnerabilities. 
For instance, the high interconnection among devices and remote access points provides entry points for the attacker that, by compromising a single node, may be able to inject malicious data into the SG network. 
Furthermore, the use of legacy systems, the inherent system complexity, and the lack of standardization render managing the large system represented by the SG very difficult, especially from a security standpoint~\cite{nafees2023smart}. 
Despite the investigation of authentication and access control mechanisms for securely collecting and managing data in SGs~\cite{6470960,9222155,en81011883,7556606}, SGs are nowadays still an easy target for cyberattacks~\cite{forbes}.

While the successful integration of AI technologies shows that SGs are revolutionary in modernizing the electricity sector, they remain one of the most vulnerable points of SGs~\cite{7926429}. 
Indeed, a few studies~\cite{9914610,10.1145/3447555.3464859,8646424} are assessing the susceptibility of AI-enabled stability prediction systems in SGs to adversarial attacks. 
The main idea behind these attacks is to inject maliciously crafted data into the SG network to fool the AI-enabled stability prediction system and hence cause faults.
Injecting malicious data into the system transforms a potential adversarial attack into a false data injection attack targeting the entire grid.  
Such attacks not only affect the stability prediction system but also impact interconnected systems relying on accurate grid data, causing broader disruptions. 
The attacker's ability to manipulate data distribution challenges grid operators who depend on accurate information for critical decisions. 
The risk escalates as manipulations may go unnoticed when the stability prediction model is compromised.
This manipulation poses a significant risk as it could obscure any genuine instability within the grid, whether caused by the attacker or other factors.
Up to now, all studies in the literature focus on state-of-the-art adversarial attacks, which however can be mitigated via state-of-the-art solutions.
However, no proposal in the literature design attacks specifically for stability prediction systems leveraging mild assumptions related to the knowledge of data and model parameters. 
This represents a fundamental need to address, as attacks on prediction systems may lead to severe malfunctioning, resulting in a lack of service and/or disruption of critical components of the infrastructure (e.g., due to overvoltage).
SGs are part of a nation's critical infrastructures and need hence to be secured against these threats.

In this paper, we introduce GAN-GRID, a novel Adversarial attack using a Generative Adversarial Network (GAN) to generate grid-like data classified as stable by an ML-based stability prediction system.
To the best of our knowledge, this is the first contribution proposing a new adversarial attack that requires minimal access to the real data and the model and demonstrates high success rates against stability prediction systems in SGs.
Given the absence of openly available code for stability prediction systems in state-of-the-art papers, we first develop and test different ML and DL models specific to stability prediction tasks, achieving up to 0.999 Accuracy. 
We then propose a novel adversarial model specifically targeting stability prediction systems.
Starting from random data, our attack leverages a GAN optimized by reinforcement learning.
When developing adversarial attacks, access to data and model specifics is crucial for creating effective adversarial samples that mislead the stability prediction system. 
Based on this consideration, we evaluate the vulnerability of these models to our attack in both a white box (i.e., access to model and data) and a grey box (i.e., access to model output) scenario, showcasing susceptibility even without access to authentic data or model details.
The resulting injected data poses serious risks as it does not trigger any alarms regarding instability within the stability prediction system.
Thus, other interconnected systems that rely on accurate grid data predictions could also be compromised.
Our contributions can be summarized as follows.
\begin{itemize}
    \item We propose a novel realistic threat model that reflects a real-world scenario of an attack on a stability prediction system that has not been discussed before in literature. 
    \item We propose \textbf{GAN-GRID}, a novel class of adversarial attacks to stability prediction systems To the best of our knowledge, we are the first to develop such attacks in this context.
    \item We propose and evaluate several stability prediction models to determine which are the most effective for stability prediction applications. Our evaluation together with our open-source code, provides a reference for future studies on stability prediction models and their security.
    \item We evaluate our system and attacks on the Electrical Grid Stability Simulated Dataset.
    We show an accuracy of up to 0.999 for our stability prediction models. Also, our attack was able to deceive the stability prediction models to classify the generated data as stable with an Attack Success Rate (ASR) of up to 0.99. Notably, it outperformed other attacks in both ASR and the level of access required to execute the attack.
    \item We make the code of our systems, attacks, and the dataset available at: \url{https://anonymous.4open.science/r/GAN_GRID-435F}. Thanks to our code, we foster research on this subject providing a common baseline for future evaluation and developments.

\end{itemize}

\section{Related Work}
In this section, we present related works on stability prediction systems and their security.
In particular, we review existing stability prediction methodologies in Section \ref{sec:stability}, while we review currently available attacks to these systems in Section \ref{sec:revAttacks}.

\subsection{AI and ML for SG Stability}\label{sec:stability}
In this context, AI has emerged as one of the most transformative and impactful technologies for the effective management of power grids and SGs~\cite{https://doi.org/10.1002/2050-7038.12706}. 
These cutting-edge AI techniques offer powerful and promising solutions for the stability analysis and control of SGs, attracting increasing interest and attention from researchers and practitioners alike~\cite{SHI2020115733}. 
For instance to predict SG stability optimized Deep Learning (DL) models analyze the Dynamic Synchronous Generator Controller (DSGC) system across diverse input values, achieving up to 99.62\% accuracy~\cite{breviglieri2021predicting}. 
Önder et al.~\cite{onder2023classification} introduced five distinct cascade methodologies, encompassing pre-processing, training, testing division, and classification stages within the stability estimation procedure for SGs. 
Bashir et al.~\cite{https://doi.org/10.1002/2050-7038.12706} utilized a range of state-of-the-art ML algorithms, 
including Support Vector Machines (SVM), K-Nearest Neighbor (KNN), Logistic Regression, Naive Bayes, Neural Networks, and Decision Tree classifiers, to forecast SG stability. Gorzałczany et al.~\cite{en13102559} tackle the challenge of transparent and precise prediction of decentralized SG control stability by leveraging a knowledge-based data-mining methodology, specifically a fuzzy rule-based classifier. Their approach utilizes multi-objective evolutionary optimization algorithms to enhance the balance between interpretability and accuracy within the classification system. An improved model is introduced in~\cite{sym15020289}, harnessing the capabilities of explainable AI and feature engineering for predicting SG stability. Notably, this study adopts a symmetrical approach by addressing the problem from both classification and regression perspectives. Dewangan et al.~\cite{DEWANGAN2022149} have presented a new and enhanced genetic algorithm (GA)-based extreme learning machine (ELM) model for forecasting the stability of SG. They explore the outcomes of this model and compare them with those of other modern AI and DL models for comprehensive analysis.
Furthermore, there is a growing emphasis on the utilization of Recurrent Neural Networks (RNNs) such as Long Short-Term Memory Network (LSTM) and Gated Recurrent Unit (GRU) in the literature~\cite{10.1145/3459104.3459160,9079864,9446196}. 
Their widespread adoption underscores their effectiveness in capturing temporal dependencies and modeling sequential data, thus enhancing the accuracy and reliability of stability prediction systems in SG environments. 
Convolutional Neural Networks (CNNs) are emerging as a popular choice in stability prediction research within SGs, evidenced by their recurrent application in the literature~\cite{AHAKONYE2024101086,8486644,SHI2020114586,10386170}. 

\subsection{Adversarial Attacks}\label{sec:revAttacks}
Ahmadian et al.~\cite{8646424} introduced a False Data Injection Attacks (FDIA) utilizing a GAN architecture. In this model, the attacker assumes the role of the generative network, while the Energy System Operator (ESO) acts as the discriminative network. By formulating an optimization problem, the attacker generates deceptive data that evades detection by the power system state estimator. Li et al.~\cite{9303013} illustrate the susceptibility of well-established ML models used for detecting energy theft to adversarial attacks. Specifically, they develop an approach for generating adversarial measurements, allowing attackers to report significantly reduced power consumption to utility companies, effectively evading detection by the ML-based energy theft detection systems. Chenet al.~\cite{8587547} endeavor to tackle security concerns surrounding ML applications within power systems. They highlight that the majority of ML algorithms currently proposed for power systems exhibit vulnerability to adversarial examples, which are inputs deliberately crafted with malicious intent. Sayghe et al.~\cite{9281719} explore the influence of adversarial examples on the detection of FDIA utilizing DL algorithms. Their study delves into the effects on Multilayer Perceptron (MLP) when subjected to two distinct adversarial attack methodologies. As discussed in the literature, ML/DL models are frequently employed as stability prediction systems, yet they are vulnerable to adversarial attacks, an issue often overlooked in previous research~\cite{HAO2022123}.
\section{System and Threat Model}
\label{sec:systhreat}
\noindent{\textbf{System Model.}}
In an operational scenario devoid of active threats targeting system disruption, the stability prediction system receives input data from the SG infrastructure, i.e., different sensors and Phasor Measurement Unit (PMU) measurements from different points across the grid. 
The stability prediction model is designed to analyze grid conditions and determine whether stability is maintained or compromised. 
Thus, this system focuses solely on stability prediction, which entails discerning whether the grid is stable or unstable (binary classification task).
To this aim, it uses ML and/or AI algorithms to discern whether, based on the current observations, the SG will be stable or not in the near future.
Before deployment, the stability prediction model undergoes training using uncorrupted data to ensure accurate and reliable predictions within the operational environment.\\

\noindent{\textbf{Threat Model.}}
The attacker's goal is to inject fraudulent data into the grid's stream, covertly aiming to manipulate the stability model's classification. 
To this aim, the attacker might exploit known or new vulnerabilities to gain remote access~\cite{chen2011lessons,sullivan2017cyber}.
The primary goal is to deceive the stability prediction system into classifying the injected packets as belonging to the stable class. 
We define two scenarios based on the attacker’s knowledge of the SG's data and of the stability prediction model.

\begin{itemize}
    \item \textit{White-box Scenario}: In this scenario, the attacker possesses comprehensive access to both the data employed in testing the model and detailed information regarding the model's architecture and parameters. 
    This advantageous position provides the attacker with ample opportunities to exploit vulnerabilities in the system. 
    By leveraging this intelligence, the attacker can meticulously craft powerful adversarial samples aimed at deceiving the model. 
    Additionally, having access to the model weights enables the adversary to fine-tune the attack parameters offline, enhancing the effectiveness and sophistication of their attacks.
    
    \item \textit{Gray-box Scenario}: In real-world contexts, scenarios where adversaries successfully infiltrate systems to compromise stability prediction models through unauthorized access to data and trained models are rare. 
    Various defense strategies outlined in the literature empower real-world systems to integrate countermeasures aimed at deterring direct breaches~\cite{8887286,9955605,efatinasab2024faultguard}. 
    It is also suggested by~\cite{10.1145/3447555.3464859} that while we shouldn't dismiss the potential for input-specific adversarial attacks, they are generally considered less plausible as attacks against SG stability assessment systems because of their access requirements to data and the internal model.
    In a more realistic scenario, termed a grey-box setting, attackers can only access the trained models' output without obtaining data from the grid or accessing the model architecture and training details. 
    However, it's crucial to note that attackers may possess knowledge of the features used by the stability prediction model for the development of adversarial attacks, which could be inferred from widely available literature or through interactions with the model itself.
    In this grey-box scenario, the attacker preemptively uses the model output to train the generator of a GAN. 
    By leveraging the stability prediction model as an oracle, the attacker can train a neural network using its feedback.
    \end{itemize}


\section{GAN-GRID: Our Proposed Adversarial Attack}
\label{sec:attacks}
We now discuss the attacks that we employ against stability prediction systems in SGs. 
In Section \ref{sec:gbox} we describe our proposed methodology to generate adversarial samples in a greybox setting.
In Section \ref{sec:wb} we then present common whitebox adversarial approaches that represent a baseline for comparison with our proposed attack model.

\subsection{GAN-GRID Model}\label{sec:gbox}

\begin{figure}[!h]
    \centering
    \includegraphics[width=.8\textwidth]{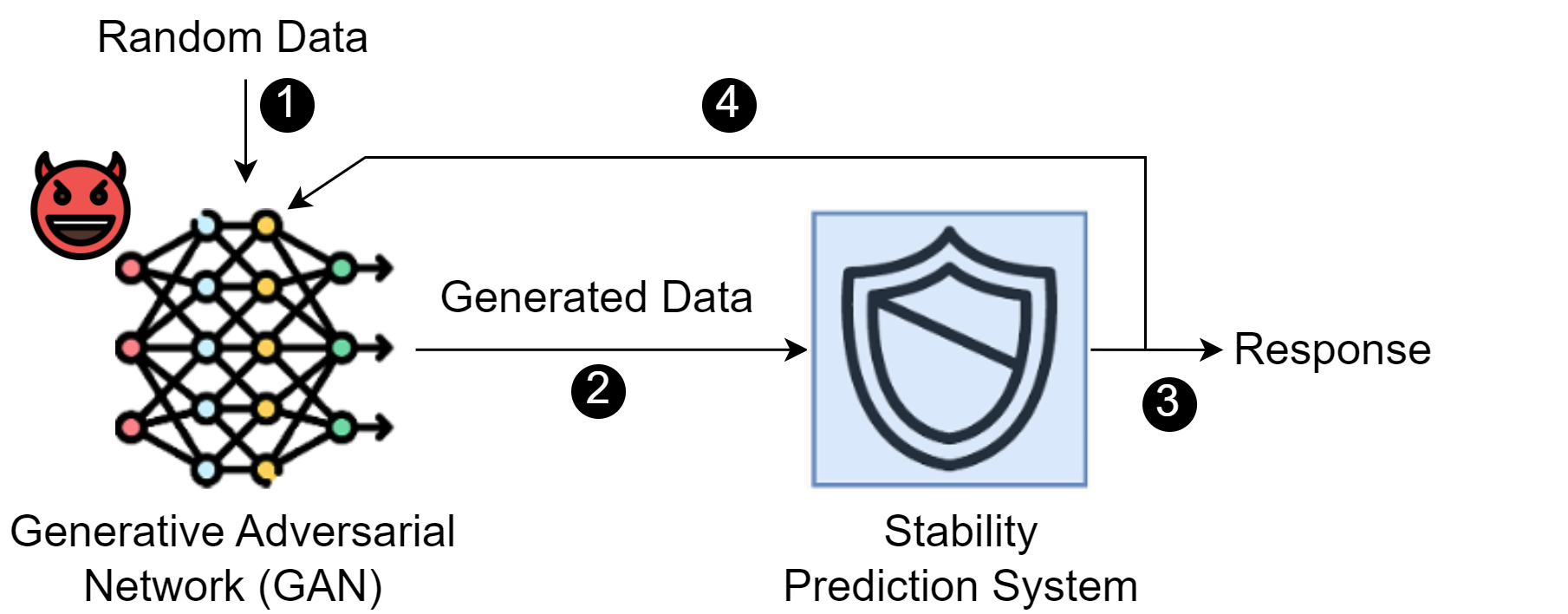}
    \caption{GAN-GRID Attack Workflow.}
    \label{fig:pipeline}
\end{figure}

In this section, we describe the workflow of GAN-GRID as depicted in Figure \ref{fig:pipeline}. 
In our scenario, the attacker gains access to the stability prediction model response without direct access to the underlying data. 
This is akin to a modified GAN training process, where the attacker utilizes the legitimate model to train the generator.

The attacker starts by providing input to the GAN randomly sampled data \ding{182}.
The output of the GAN network is then distributed in the SG grid network \ding{183}.
We optimize our generator model in conventional GAN training to outsmart a fixed discriminator, represented by the stability prediction system, rather than training both components iteratively. 
By leveraging the stability prediction model as an oracle \ding{184}, the attacker trains a neural network capable of generating fraudulent samples, even from random data. 
Our generative network leverages discriminator feedback \ding{185}, provided by the stability prediction system's output, for optimization and loss computation. 
This feedback guides the generator model in producing fraudulent data that can trick the stability prediction system.
To address the challenges of convergence and navigating local minima in the large search space, we use reinforcement learning to improve the generator's learning procedure. 
This strategic choice allows for more efficient exploration of the search space and adaptation in response to feedback and rewards received during the learning process. 
By employing exploration and exploitation strategies, the generator can strike a balance between trying new approaches and leveraging existing knowledge to identify promising search spaces.
Through the integration of reinforcement learning techniques, our approach transcends the limitations typically associated with traditional optimization methods.

The reinforcement learning process involves several key parameters, including the maximum episode length, discount factor denoted as $\gamma$, the number of episodes, and the learning rate represented by $\alpha$. 
These parameters govern the update mechanism for the generator's latent input using reinforcement learning.
The training loop operates across episodes, where each episode begins by initializing the latent input parameters and the episode reward. 
Within each episode, the generator generates a sample based on the latent input. 
This generated sample undergoes evaluation by the stability prediction model, which provides predictions against randomly generated target labels for comparison.
The reward is computed as the mean accuracy of the predictions matching the targets. 
To update the latent input using reinforcement learning, the temporal difference error ($td_{error}$) is calculated as the difference between the reward and the cumulative episode reward. 
The reward reflects the agent's performance in an episode, offering immediate feedback on its decisions. 
Conversely, the cumulative episode reward signifies the total reward gathered throughout an entire episode, bounded by the maximum number of steps or actions allowed in the reinforcement learning process. 
By calculating the $td_{error}$ as the difference between the reward and the cumulative episode reward, we capture the discrepancy between the immediate feedback received and the overall performance over an extended period. 
Subsequently, the latent input is updated by incorporating a scaled noise term to introduce randomness and facilitate exploration. 
The scaling factor for the noise term is determined by $\alpha$, $td_{error}$, and the $\gamma$ factor raised to the power of the current step. 
Mathematically, the update equation for the latent input is expressed as:

\begin{equation}
    latent\_input = \alpha \cdot td_{error} \cdot \gamma^{\,step} \cdot latent\_input.
\end{equation}

This scaling factor influences the magnitude of the noise added to the latent input, potentially increasing or decreasing the level of exploration based on the $td_{error}$’s magnitude. 

Scaling the noise with $td_{error}$ enables dynamic exploration adjustment during training. 
Higher $td_{error}$ yields larger scaling factors, increasing exploration and randomness in latent input updates. 
Conversely, lower $td_{error}$ results in smaller scaling factors, decreasing exploration and increasing exploitation as the agent refines estimates and converges towards better solutions, reducing randomness in latent input updates. 
This mechanism allows the generator to adapt its latent input based on the reward signal, facilitating the exploration of diverse latent space regions.
As the agent learns from experience, the future rewards' impact on the scaling factor diminishes, allowing the agent to prioritize immediate feedback for policy optimization. 
After each episode, the generator updates using the final latent input.
The stability prediction model assesses the generator's output, generating a target label tensor for loss calculation. 
Binary cross-entropy loss computes the generator's loss, and parameters are updated via backward propagation. 
Upon completing the specified number of episodes, the trained generator is returned, capable of producing deceptive data without knowing the real data distribution.
This updating mechanism enables the generator to adapt its latent input according to the received reward signal, allowing it to explore diverse regions within the latent space. 
As the agent gains more experience and learns from previous steps, the influence of future rewards on the scaling factor decreases, allowing the agent to focus more on optimizing its policy based on immediate feedback. 
Following each episode, the generator undergoes an update using the final latent input.
Once the designated number of episodes is completed, the trained generator is returned, equipped with the capacity to generate deceptive data effectively even without a glance at real data distribution. 

\subsection{Reference Whitebox Attacks}\label{sec:wb}
In a white-box threat model, the adversary is equipped with complete knowledge of both the data utilized and the trained model itself. 
Consequently, we undertake an examination of notable adversarial attacks to unveil vulnerabilities inherent in these models. 
Notice that this setting represents the most advantageous one for the attacker. 
Consequently, since this has been widely studied in the literature, we leverage well-studied and understood attacks as a reference to evaluate GAN-GRID that leverages a less advantageous graybox setting.
Our attention is directed toward specific attacks that have been emphasized in the literature due to their significance and effectiveness in uncovering weaknesses within ML models. 
However, it is important to note that many well-known attacks have not been tested or implemented in libraries for binary classification problems. 
This constraint posed challenges in identifying and selecting appropriate attack methodologies.
\begin{itemize}
  \item \textit{Fast Gradient Sign Method (FGSM):} FGSM efficiently generates adversarial examples by leveraging the gradient sign of the loss function. Renowned for its computational efficiency, FGSM serves as a fundamental benchmark for assessing model robustness~\cite{goodfellow2015explaining}.
  
  \item \textit{Basic Iterative Method (BIM):} BIM builds upon FGSM by iteratively applying small perturbations at each step, thereby enhancing the attack's potency. This iterative approach offers insights into the cumulative effects of perturbations, shedding light on nuanced aspects of model robustness~\cite{kurakin2017adversarial}.
  
  \item \textit{Projected Gradient Descent (PGD):} PGD adopts an iterative optimization strategy similar to BIM, but distinguishes itself by incorporating a projection step to confine perturbations within a predefined constraint set. This distinctive feature enables PGD to craft highly potent adversarial examples, facilitating thorough examination of model robustness under rigorous conditions.~\cite{madry2019deep}. 
 
  \item \textit{Random noise:} This custom implementation of random noise attack strategy utilizes a method of introducing random noise to generate adversarial instances aimed at undermining our models. The attack introduces random perturbations drawn from a normal distribution to the original samples. Each input sample undergoes multiple iterations of perturbation, guided by the user-defined epsilon parameter, which controls the magnitude of the perturbation. Following perturbation, the samples are subjected to the models classification process. If the resulting accuracy is lower than the original predictions, signifying successful deception, the perturbed sample replaces the original in the set of adversarial examples. This iterative process continues until either a successful adversarial instance is identified or the maximum number of perturbation attempts, specified by the number of samples parameter, is exhausted. We opted for a sample size of 50 to minimize computational burden.
\end{itemize}

\section{Grid Stability Prediction}
\label{sec:delamain}
In this section, we thoroughly explore models developed specifically for stability prediction. 
Despite the presence of a vast literature that proposes models for stability prediction, we explore new models for stability prediction in response to a critical concern. 
While some models in the literature may perform satisfactorily, their reproducibility is a significant limitation. 
Indeed, the lack of sufficient information about the model architecture and hyperparameters or the lack of their open-source code prevents accurate replication of these models. 
Therefore, we resort to creating state-of-the-art-based stability prediction models to test the effectiveness of our devised attack. 
To ensure a thorough and complete analysis, we employ both classical ML (Section \ref{sec:ml}) and DL (Section \ref{sec:dl}) models. 
This dual approach helps us understand their performance and vulnerability to attacks comprehensively, drawing robust conclusions about stability prediction efficacy and security against potential threats.
\subsection{ML Model Design}\label{sec:ml}
In our ML model implementation, we consider classical ML algorithms such as Decision Trees, Extra Trees, XGBoost, KNN, Light Gradient-Boosting Machine (LGBM), and Random Forest. 
After thorough training and comparison experiments with other algorithms (see Section \ref{sec:baseline} for details), we selected the XGBoost architecture. 
Following hyper-parameter tuning, XGBoost emerged as the optimal choice for our stability prediction system due to its superior performance and lightweight nature. 
This efficiency ensures swift data processing and model evaluation, making it well-suited for real-time prediction tasks and enhancing the responsiveness and reliability of our system.

\subsection{DL Model Design}\label{sec:dl}
To ensure practicality and efficiency, we engineered our DL stability prediction model to be streamlined, minimizing computational demands while maximizing effectiveness. 
This design philosophy aligns with our goal of creating a robust yet resource-efficient system.
Our stability prediction model employs a one-layer Bi-directional LSTM architecture with 220 neurons to capture temporal dependencies in both forward and backward directions within the time sequence. 
To prevent overfitting, we introduced a dropout layer with a 0.5 dropout rate during training. 
Following the dropout layer, the LSTM layer's output passes through a Linear layer with 440 neurons, activating an element-wise sigmoid function. 
The deliberate choice of LSTMs aims to capture potential causal relationships between data points.
For model optimization, we use the Binary Cross-Entropy loss function, a standard metric for binary classification tasks. 
The training process utilizes the Adam optimizer with a learning rate of $1 \times 10^{-3}$ for efficient gradient descent. 
We structure training iterations into 10 epochs to balance duration and performance.

\section{Evaluation}
We now delve into the evaluation of the attack and baseline stability prediction systems. As metrics, we use accuracy and F1 score to evaluate the models and Attack Success Rate (ASR) to evaluate attacks, defined as:

\begin{equation}
    Accuracy = \frac{TP + TN}{TP + FP + TN + FN},
\end{equation}

\begin{equation}
    F1 = \frac{2TP}{2TP + FP + FN},
\end{equation}

\begin{equation}
    ASR = \frac{\text{\# malicious batches fooling the stability prediction}}{\text{\# malicious batches sent}}.
\end{equation}
\label{sec:evaluation}

\subsection{Dataset}
\begin{figure}[!h]
     \centering
     
  \begin{subfigure}[]{0.3\textwidth}
     \centering
     \includegraphics[width=\textwidth]
        {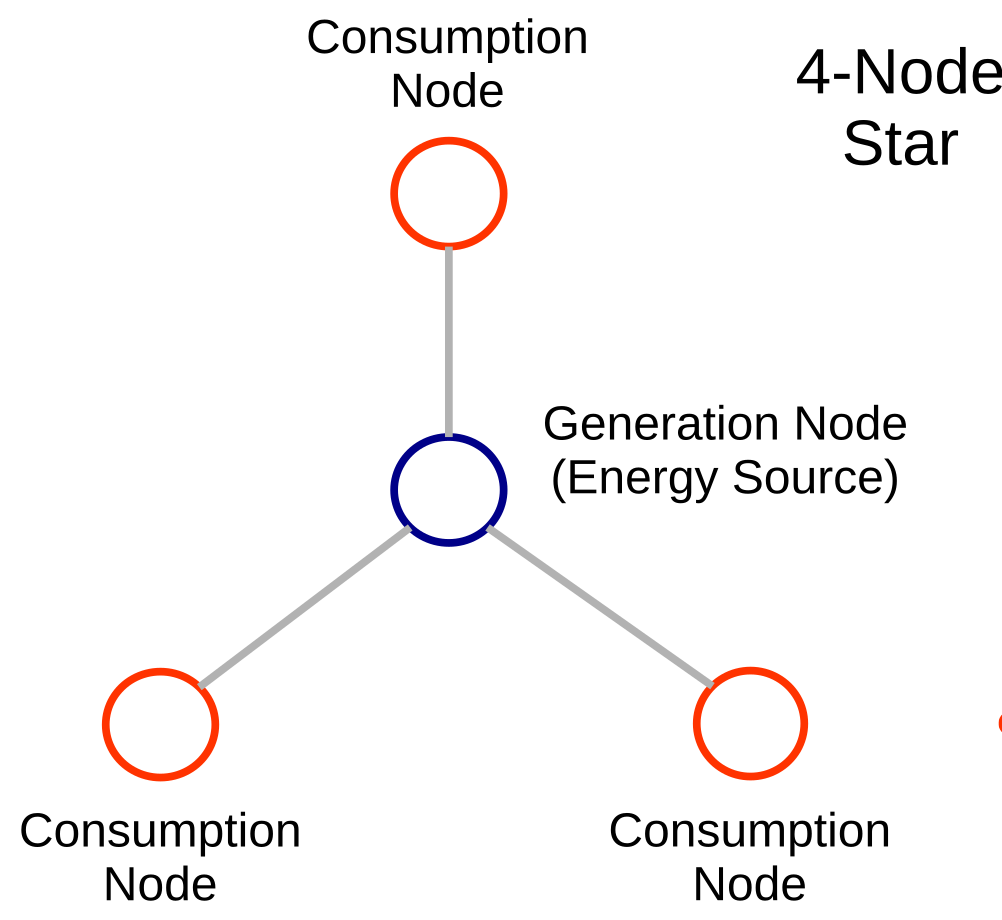}
     \caption{Topology: 4-node star architecture}
        \label{fig:star_architecture_1}
 \end{subfigure}
\hfill
 \begin{subfigure}[]{0.3\textwidth}
     \centering
     \includegraphics[width=\textwidth]
        {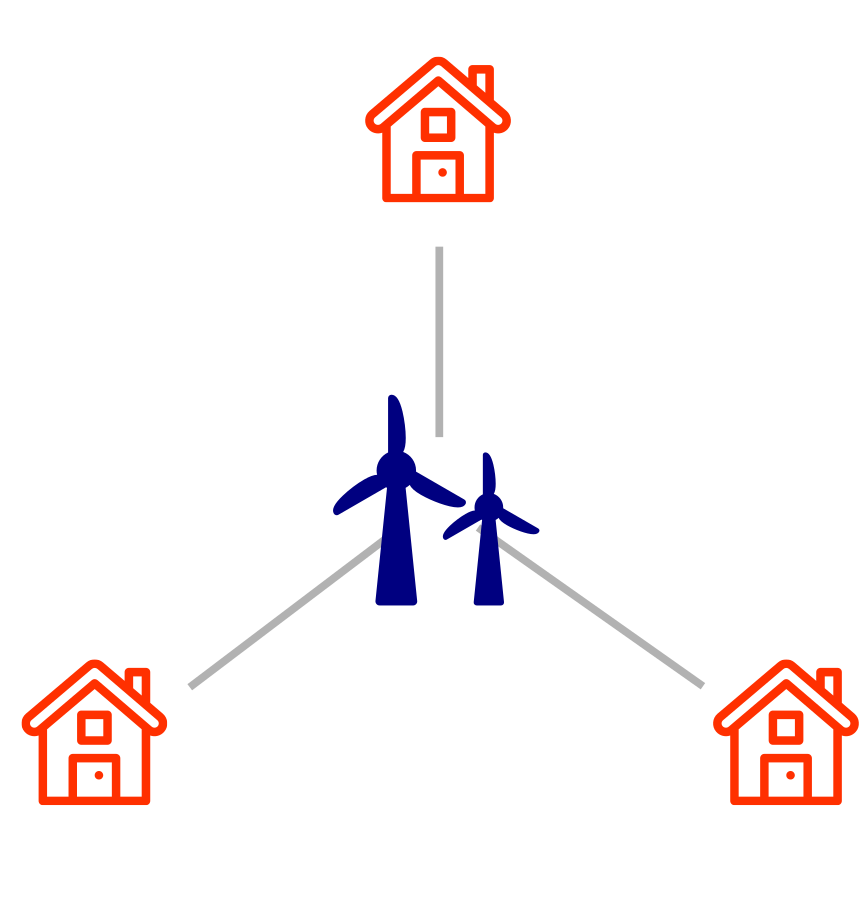}
     \caption{Real-world example of the topology}
        \label{fig:star_architecture_2}
 \end{subfigure}
  \caption{Laboratory setup for the real attack experimentation.}
    \label{fig:arch}
 \end{figure}

The dataset utilized for evaluating our systems originates from an augmented version of the \emph{Electrical Grid Stability Simulated Dataset} obtained from the University of California (UCI) Machine Learning Repository~\cite{misc_electrical_grid_stability_simulated_data__471}. 
Initially containing 10,000 samples, this dataset contains simulation outcomes regarding grid stability for a reference 4-node star network, as depicted in Fig~\ref{fig:star_architecture_1}. 
Also a real-world example of such architecture can be seen in Fig~\ref{fig:star_architecture_2}. 
By augmentation, the dataset expanded to 60,000 samples, leveraging the grid's inherent symmetry and increasing the dataset sixfold. 
It comprises 12 primary predictive features and two dependent variables, offering insights into grid stability dynamics. 
To manage the dataset effectively, we used a robust windowing technique, segmenting it into predefined-size segments. 
Each window was created iteratively by traversing the data with a step size equal to half of the window size, set at 16  for our dataset. 
Additionally, we partitioned the dataset into training (75\%), validation (5\%), and test (20\%) subsets. 
Preprocessing steps focused on normalization to prepare the dataset for prediction models effectively.

\subsection{Baseline Evaluation}\label{sec:baseline}
During the evaluation phase, we assess the performance of our stability prediction systems. 
We first utilize the training data to train both ML and LSTM models. 
Subsequently, we evaluate the efficacy of our stability prediction system on the test set. 
The results of our evaluation are noteworthy. 
The Best ML model, i.e., XGBoost, 
achieves a mean accuracy of 0.994±0.001, while the DL model demonstrates even higher accuracy, reaching 0.999±0.001 for the stability prediction task. A comprehensive presentation of results is provided in Table~\ref{tab1}.
\begin{table}[htbp]
\caption{ML and DL Models Performance Metrics}
\begin{center}
\begin{tabular}{l|c|c}
\hline
        \multirow{2}{1cm}{\textbf{Model}} & \multicolumn{2}{c}{\textbf{Performance Metrics}} \\ \cline{2-3}
        & Accuracy & F1 Score \\
        \hline
LSTM & 0.999 & 0.999\\

XGBoost & 0.994 & 0.994 \\

LGBM & 0.97 & 0.97\\

Decision Tree & 0.974 & 0.974\\

Extra Trees & 0.991 & 0.991\\

KNN & 0.875 & 0.874\\

Random Forest & 0.988 & 0.988\\
\hline
\end{tabular}
\label{tab1}
\end{center}
\end{table}

\subsubsection{Feature Importance}
\label{sec:shap}
To discern the most influential features employed by both DL and ML models, we employ Explainable Artificial Intelligence (XAI) techniques. 
Specifically, we leverage SHapley Additive exPlanations (SHAP)~\cite{10.5555/3295222.3295230}, recognized for its model-agnostic nature and robust interpretability.
SHAP allows us to quantify the contribution of each feature to the model's predictions, offering insights into the underlying decision-making process. 
We use SHAP Gradient Explainer for interpreting the LSTM model and SHAP Tree Explainer for the XGBoost model, with results depicted in Fig.~\ref{fig1} and Fig.~\ref{fig2}.
The analysis indicates varying feature importance between XGBoost and LSTM models. 
In XGBoost, participant reaction time (tau[x]) is primary, followed by price elasticity coefficients (gamma). 
Nominal power consumption or production features (p[x]) have less impact. 
In contrast, the LSTM model prioritizes price elasticity coefficients and then participant reaction time. 
However, both models consider nominal power consumption or production features less critical in decision-making processes. 
This observation aligns with findings from the literature, where Erdem et al.~\cite{10.1007/978-3-031-04112-9_24} utilized Layer-Wise Relevance Propagation (LRP) to determine relevance scores for each input, thereby confirming the diminished importance of nominal power consumption or production features in decision-making processes.

\begin{figure}[!h]
     \centering
     
  \begin{subfigure}[]{0.48\textwidth}
     \centering
      \includegraphics[width=\textwidth]
    {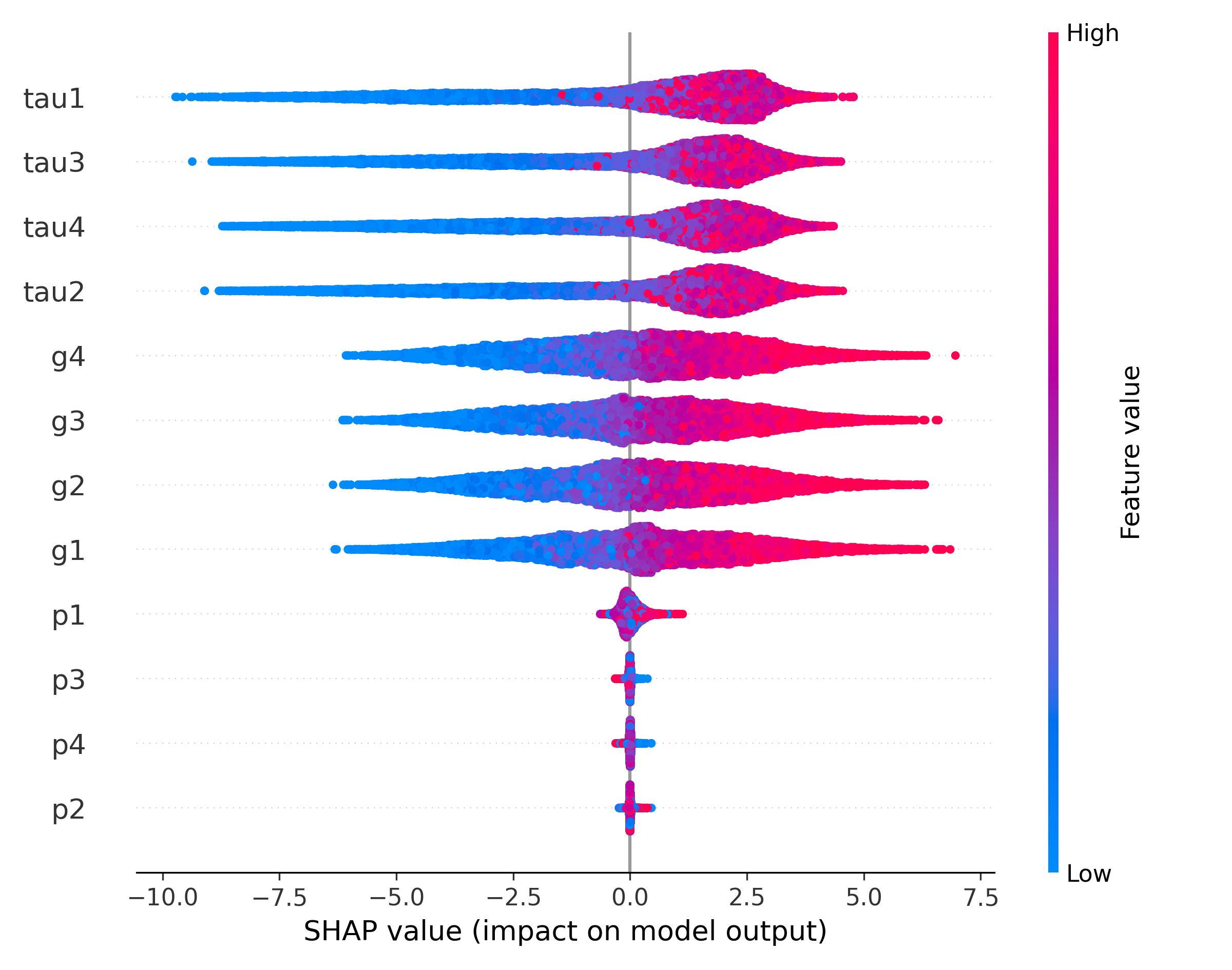}
    \caption{Shap values for XGBoost.}
    \label{fig1}
 \end{subfigure}
\hfill
 \begin{subfigure}[]{0.48\textwidth}
     \centering
     \includegraphics[width=\textwidth]
    {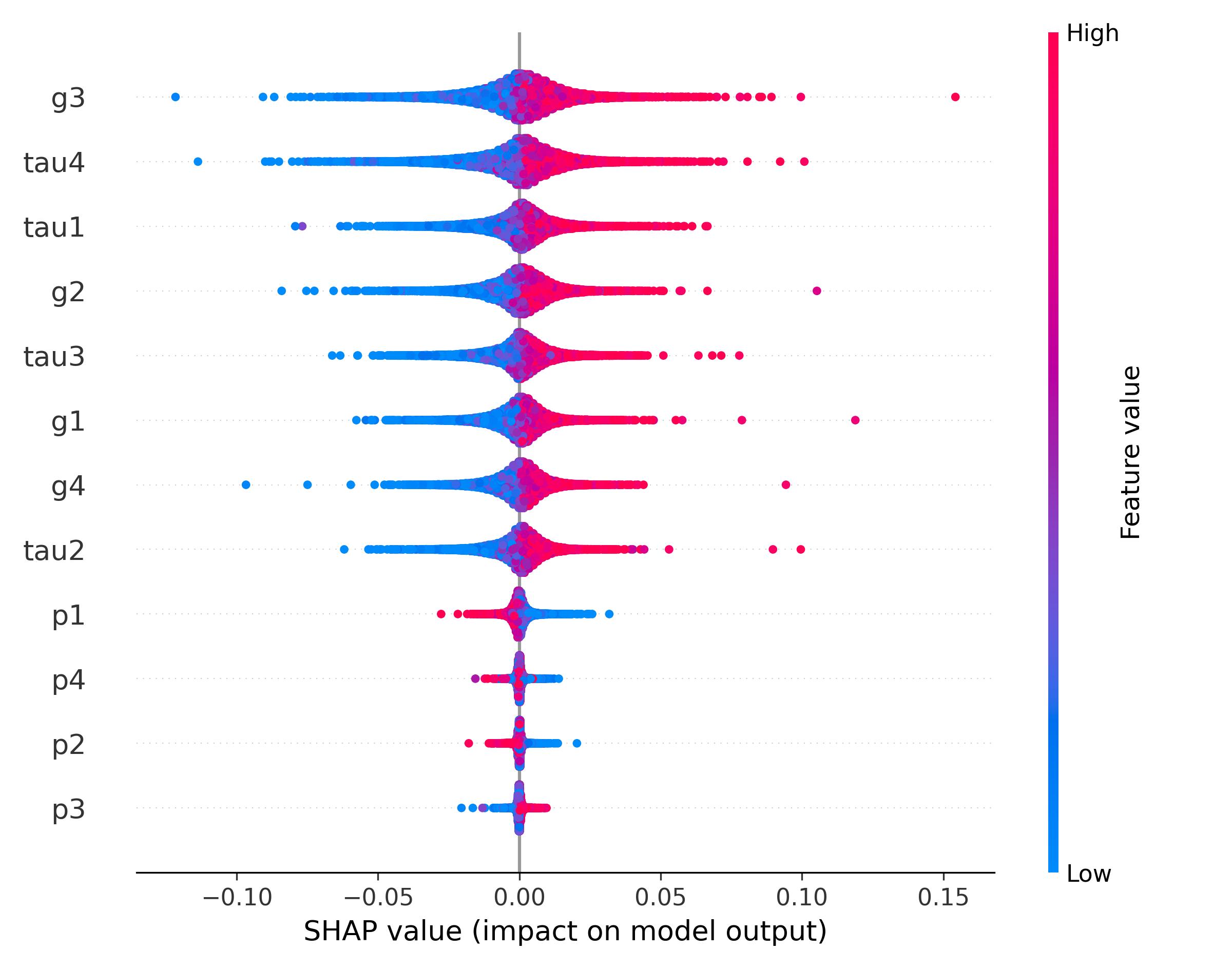}
     \caption{Shap values for LSTM.}
         \label{fig2}
 \end{subfigure}
  \caption{Shap values.}
    \label{fig:shap}
 \end{figure}
 
\subsection{Attack Evaluation}
We proceed to assess our attacks against the stability prediction models, dividing the evaluation into the scenarios outlined in the threat model in Section~\ref{sec:systhreat}: white-box attacks and the GAN-GRID attack.

\begin{figure}[!h]
     \centering
     
  \begin{subfigure}[]{0.48\textwidth}
     \centering
      \includegraphics[width=\textwidth]
    {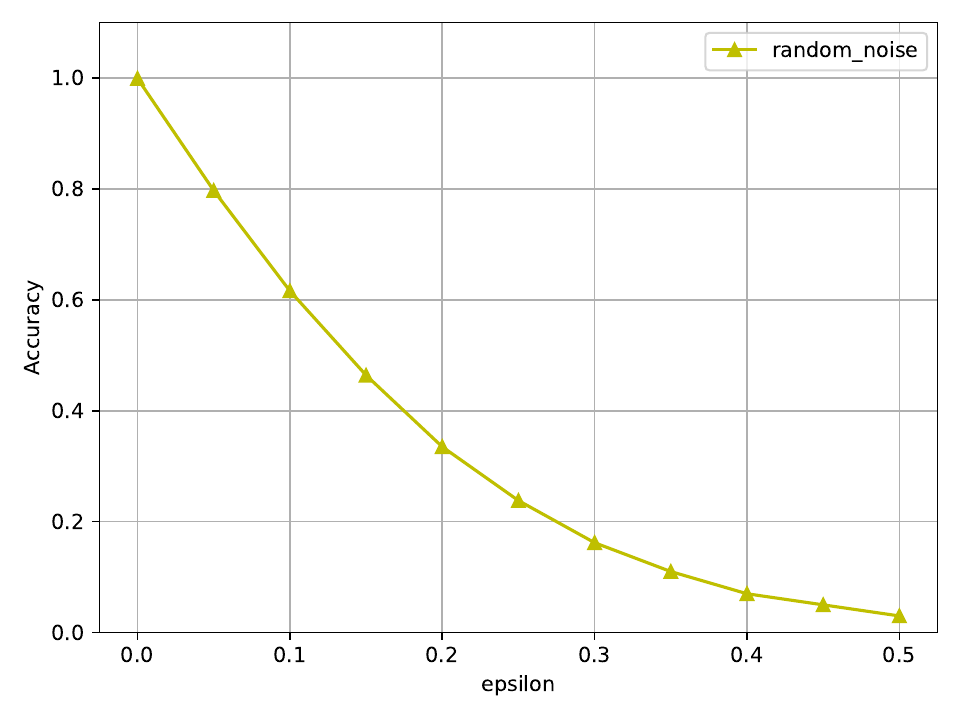}
    \caption{Model accuracy vs epsilon for XGBoost.}
    \label{fig1}
 \end{subfigure}
\hfill
 \begin{subfigure}[]{0.48\textwidth}
     \centering
     \includegraphics[width=\textwidth]
    {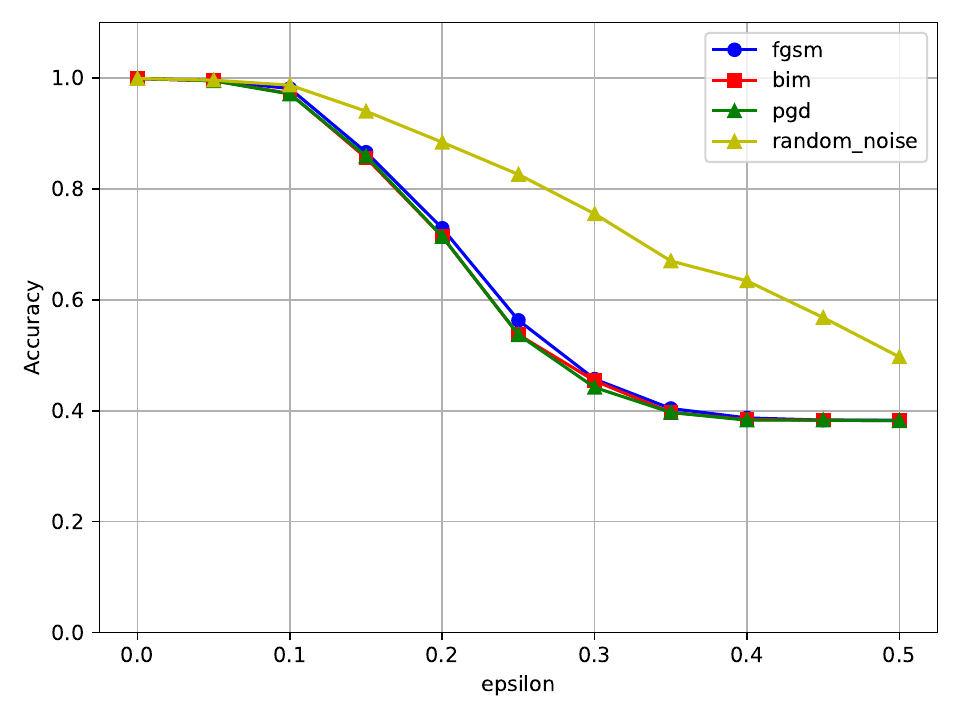}
     \caption{Model accuracy vs epsilon for LSTM.}
         \label{fig2}
 \end{subfigure}
  \caption{Model’s accuracy at varying epsilon values on the white-box attacks.}
    \label{fig:accuracy}
 \end{figure}

\paragraph{\textbf{White-box Evaluation.}}
\label{subsec:WB}
In this section, we thoroughly assess the effectiveness of white-box attacks, as detailed in Section~\ref{sec:wb}. 
To execute these attacks, we utilize the Adversarial Robustness Toolbox (ART) library~\cite{nicolae2018adversarial}, probing the baseline systems to evaluate the susceptibility of our models without incorporating any countermeasures or defenses. 
In classical ML models, characterized by non-differentiable architectures such as decision trees or ensemble methods, applying white-box adversarial attacks like FGSM, BIM, and PGD is not straightforward due to the absence of easily obtainable gradients. 
Unlike DL models, which readily provide gradients, classical ML models often lack this accessibility, rendering the application of gradient-based attacks impractical or challenging. 
This challenge extends beyond just accessibility; it also pertains to fundamental differences in architecture and the methods employed in classical ML compared to DL. 
These classical ML techniques often rely on discrete decisions and non-linear transformations, making the computation and propagation of gradients inherently difficult. 
Additionally, the library implementations of these attacks do not offer built-in support for ML classifiers. 
As a result, we do not employ these three attacks against our classical ML model. 
Instead, we utilize our proposed random noise-based attack tailored for XGBoost, to explore potential vulnerabilities and assess robustness. 
The attacks are conducted with varying epsilon values, representing the strength of each attack and the extent of perturbation introduced. 
Specifically, we explore epsilon values ranging from 0.05 to 0.50. 
The outcomes of these attacks across different models are visually depicted in Fig.~\ref{fig:accuracy}. 
The XGBoost model is more susceptible to the same random noise attack compared to the LSTM model.
Moreover, it is noteworthy that the FGSM, BIM, and PGD methods exhibit nearly identical performance, surpassing that of random noise. 
Also, increasing the epsilon value beyond 0.5 does not provide any significant advantage.

\paragraph{\textbf{GAN-GRID Evaluation.}}
\label{subsec:GB}
In our attack evaluation, we utilize a generator model optimized through reinforcement learning, leveraging the output of our stability prediction systems as surrogate data. 
Our aim is to generate data classified as stable by the prediction system, without access to actual data or model architecture and training details.
We train the generator against both XGBoost and LSTM models, with negligible training time per episode, even on CPU (1s). 
After training, we synthesize data from noise using the generator, matching the number of batches in the test set. 
We subsequently evaluate this generated data against the stability prediction models.
Results show an ASR of 0.99 ± 0.01 percent for the attack against both models. 
This highlights the vulnerability of these models to our attack, as our generator can converge to a data distribution classified as stable without access to real data.
During our experiments, we conducted multiple training iterations with the generator to determine the mean convergence episode and the required time and number of data batches for classification by the surrogate model, ensuring generator convergence.
For the LSTM model, convergence typically occurs after 15 episodes of training, requiring approximately 60 batches of data to be sent for classification. 
This process takes roughly 16 minutes. 
With the XGBoost model, convergence is achieved after about 5 episodes of training, necessitating around 20 batches of data and taking approximately 6 minutes.
These results underscore stability prediction models' vulnerability to sophisticated attacks, even with limited access to data or models, emphasizing the need for enhanced robustness and security in critical systems.
The DL model takes longer than the ML model to process. 
In our simulations, data collection for the stability prediction model happens every 16 seconds, with the model requiring the same amount of time to receive data and generate predictions.

\begin{figure}[!h]
\centering
\includegraphics[width=.4\textwidth]
{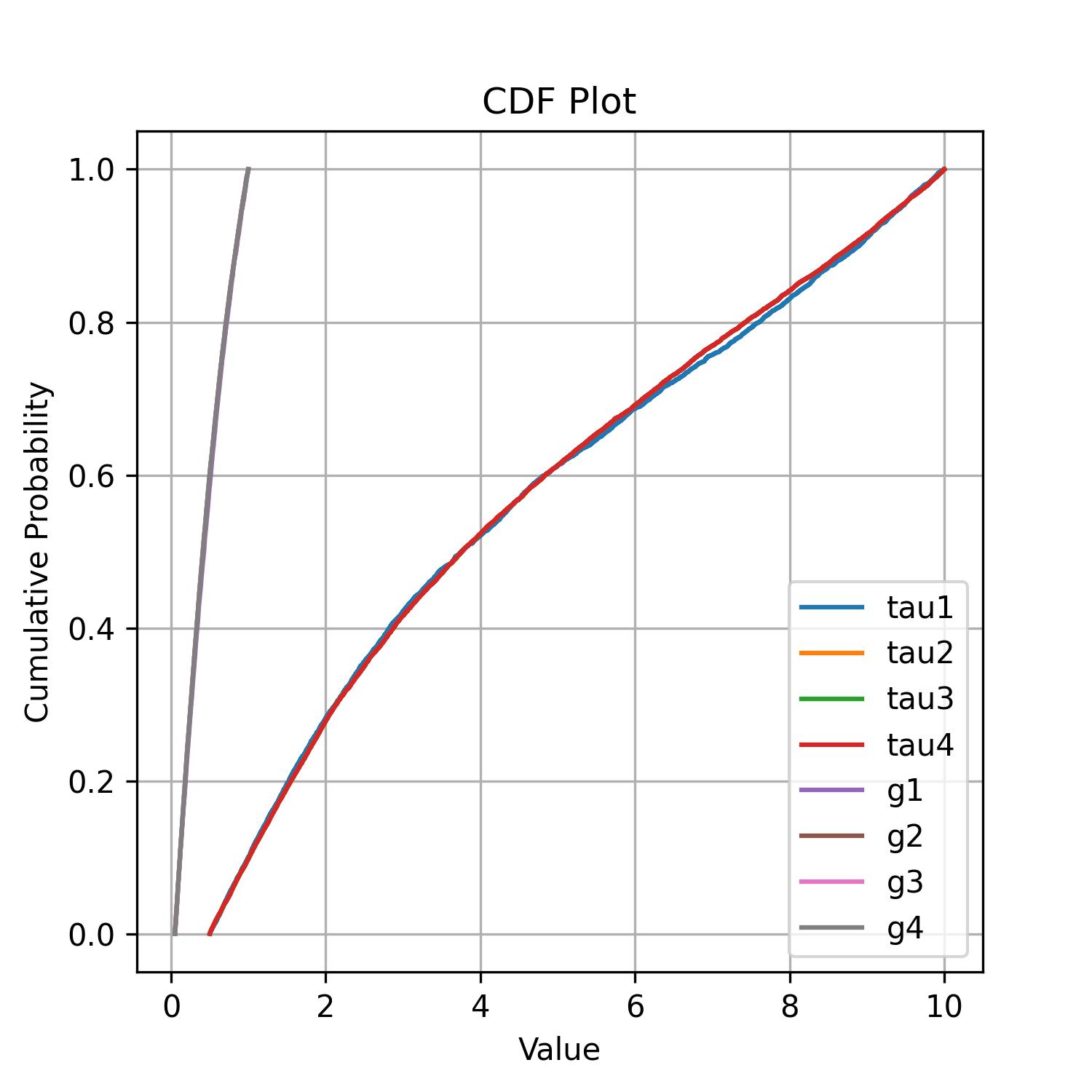}
\caption{Cumulative distribution of Real data}
\label{fig3}
\end{figure}

\begin{figure}[!htbp]
\centering
\includegraphics[width=0.4\textwidth]{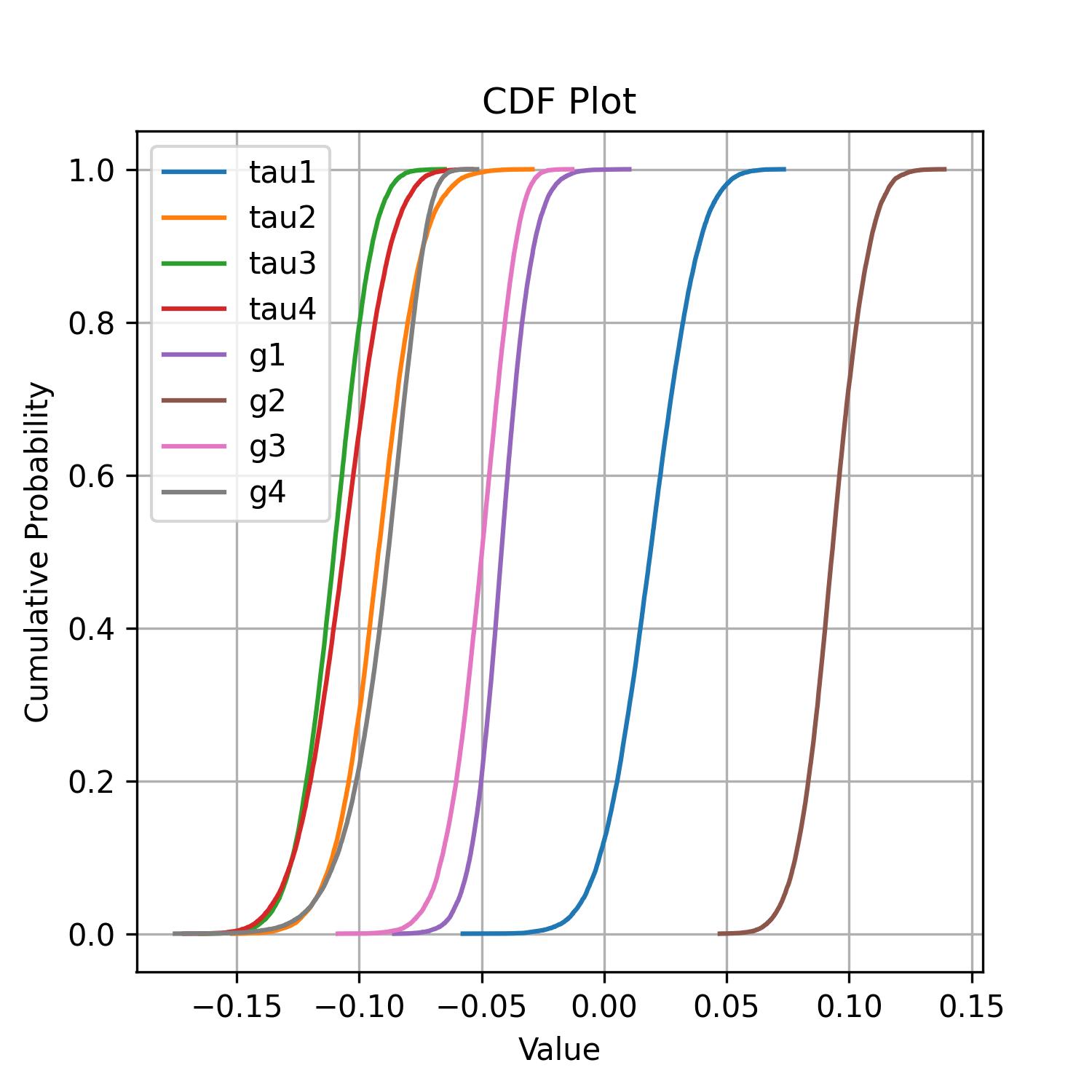}
\caption{Cumulative distribution of data generated for the ML model}
\label{fig4}
\end{figure}

\begin{figure}[!htbp]
\centering
\includegraphics[width=0.4\textwidth]{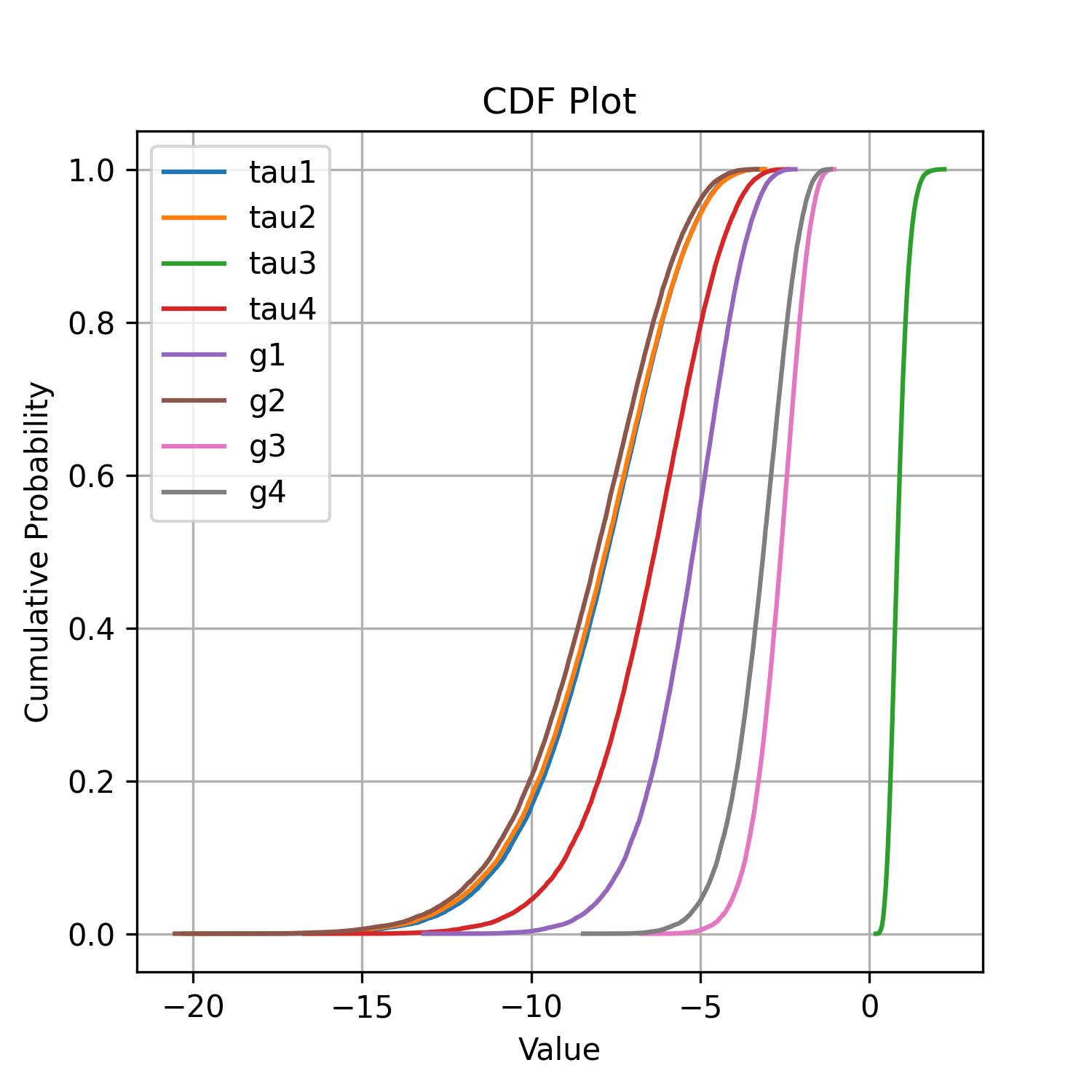}
\caption{Cumulative distribution of data generated for the DL model}
\label{fig5}
\end{figure}

In light of our discussion regarding the potential manipulation of authentic data and sensor values by malicious actors, we undertake an analysis to investigate the ramifications of the grid infrastructure. 
Our objective is to shed light on the capacity for manipulative actions to introduce distortions that could exacerbate existing grid issues while evading detection due to compromised stability systems.
The outcomes, depicted in Figs.~\ref{fig3}, ~\ref{fig4}, and~\ref{fig5}, reveal a significant discrepancy in the distribution patterns of relevant features (according to SHAP analysis in Section~\ref{sec:shap}), leaning towards smaller values compared to authentic data. 
These changes have the potential to cause significant problems within the grid infrastructure. 
Skewed distributions of relevant features towards smaller values can trigger operational challenges within the grid. 
For instance, such skewness might lead to underestimation of power demand, causing inadequate resource allocation and grid instability during peak demand periods. 
This situation can also lead to overvoltage, frequency deviations, and heightened stress on grid components, potentially resulting in equipment failures, service disruptions, and compromised grid reliability.

\paragraph{Summary.}
\label{subsec:ss}
The table~\ref{tab:evaluation1} summarizes the success rates of the outlined attacks. It is evident that white-box attacks demand extensive access to both the model and data, as discussed in our threat model in Section~\ref{sec:systhreat}. However, this scenario is often not feasible in real-world settings. On the contrary, the GAN-GRID attack merely requires access to the model's output, significantly reducing the required level of access. Moreover, in terms of 
ASR, the GAN-GRID outperforms all other attacks. Additionally, we can estimate the potential time required for an attacker to employ the GAN-GRID attack in a real scenario, further highlighting its efficiency and effectiveness.

\begin{table}[htbp]
    \centering
    \caption{Comparison of Model Performance Under Adversarial Attacks ($\epsilon=0.5$)}
    \begin{tabular}{l|c|c|c|c|c|c}
        \hline
        \multirow{2}{*}{\textbf{Model}} & \multicolumn{6}{c}{\textbf{Accuracy}} \\ \cline{2-7}
        & Baseline & GAN-GRID & FGSM & BIM & PGD & Random noise \\
        \hline
        \multirow{1}{*}{LSTM} & 0.999 & 0.01 & 0.383 & 0.382 & 0.381 & 0.497 \\
        \hline
        \multirow{1}{*}{XGBoost} & 0.994 & 0.01 & - & - & - & 0.038 \\
        \hline
    \end{tabular} 
    \label{tab:evaluation1}
\end{table}

     


\section{Conclusions}
Our study emphasizes the critical need to strengthen SG security mechanisms to defend against adversarial manipulation and maintain system stability and reliability. Using advanced ML algorithms, including XGBoost and LSTM-based DL models, we explore stability prediction using the Electrical Grid Stability Simulated dataset. Through rigorous experimentation, we achieved high predictive performance. However, our findings reveal the vulnerability of SG stability prediction systems to our novel attack, even with limited information, achieving an ASR of 0.99 outperforming other attack methods. We also demonstrated that by injecting the data generated by our attack, adversary can exacerbate grid issues without triggering alarms in compromised stability prediction systems. These results underscore the importance of enhancing resilience against cyberattacks in SG environments to ensure the ongoing integrity and efficiency of modernized electricity networks.\
\paragraph{Future Work.}
In future research, there is potential to refine the GAN-GRID attack to improve its effectiveness and success rate while reducing deployment time. This could entail exploring various generator architectures, optimization techniques, and injection strategies to optimize the attack process.
Furthermore, a primary focus will be on developing defenses against GAN-GRID attacks and investigating potential countermeasures. Additionally, examining poisoning attacks could offer valuable insights into the resilience of stability prediction systems. By establishing a new system and threat model that accounts for this type of attack, we aim to identify vulnerabilities within the models and strengthen their security posture.
Moreover, in addition to addressing GAN-GRID attacks in stability prediction systems, future research may entail evaluating the impact of these attacks on other ML-based systems, such as fault prediction systems in SGs.

%
%

%
%
%
\bibliographystyle{splncs04}
%
\bibliography{bibliography}




\end{document}